\documentclass[prl,amsmath, twocolumn, floatfix,showpacs]{revtex4}
\usepackage{rotating}

\newcommand{\lo}{\langle} \newcommand{\rc}{\rangle}
\newcommand{\ud}{{\mathrm d}}
\begin{document}

\title{Reaction rate calculation by parallel path swapping}

\author{Titus S. van Erp} \affiliation{Centrum voor Oppervlaktechemie
en Katalyse, K.U. Leuven, Kasteelpark Arenberg 23, B-3001 Leuven,
Belgium}
\begin{abstract}
The efficiency of path sampling simulations can be improved considerably using the approach of path swapping.  For this purpose, we have devised a new algorithmic procedure based on the transition interface sampling technique.  In the same spirit of parallel tempering, paths between different ensembles are swapped, but the role of temperature is here played by the interface position.  We have tested the method on the denaturation transition of DNA using the Peyrard-Bishop-Dauxois model. We find that the new algorithm gives a reduction of the computational cost by a factor 20.  
\end{abstract}

\pacs{02.70.-c, 05.20.Gg, 82.20.Pm}
\maketitle
Path sampling has become an important tool to study rare events that 
are inaccessible for straightforward molecular dynamics (MD). 
Whereas the original path sampling approach~\cite{TPS98_2} used a 
Monte Carlo sampling of trajectories with fixed lengths, 
the efficiency has been improved considerably by the introduction of the new
transition interface sampling (TIS) technique~\cite{ErpMoBol2003} that
allows flexible path lengths. Besides a reduction in the required MD steps,
it also yields a faster convergence by counting only effective crossing events.
The TIS method has been applied to various systems ranging from protein 
folding~\cite{bolhuisPNAS} to nucleation~\cite{moroni05}. 
Similar to the reactive flux (RF) approach~\cite{FrenkelSmit}, the TIS methods
allow to determine rate constants in terms of microscopic 
properties that do not sensitively depend of 
the choice of reaction coordinate (RC) and stable state definitions. 
However, whilst the efficiency of the RF methods drops dramatically
whenever the RC does not capture the exact transition mechanism, the TIS
efficiency is relatively insensitive to the 'quality' of the RC~\cite{van06}.
This is an important advantage in high-dimensional complex systems
where good  RCs can be extremely difficult to find. 
TIS has also initiated the development of
some new algorithms such as the partial path TIS (PPTIS)~\cite{MoBolErp2004}
and forward flux sampling (FFS)~\cite{FFS}. PPTIS  uses a Markovian
approximation to reduce the path length even further. FFS
was especially developed to deal with stochastic non-equilibrium systems.
Similar to RF,   
the efficiency of these methods is more sensitive to the RC~\cite{van06}. 
The advantageous scaling of TIS relies partly on the fact that it is an 
importance sampling on the
dynamical factor. 
When this factor is low,
direct evaluation as in RF becomes prohibitive.
Secondly, due to the global character of trajectories, hysteresis  effects
are less likely to occur  in path space than in phase space~\cite{van06}.
Finally, the non-locality of the shooting
move might in principle allow the sampling of multiple  
reaction channels. 
However, if the reaction channels are very distinct, it might still
take a long time for the shooting move to find these channels. In this letter,
we introduce an additional technique based on replica-exchange 
methods~\cite{marinari92}
to address this problem and test 
this method on the denaturation dynamics
of the Peyrard-Bishop-Dauxois (PBD)~\cite{PBD} 
model for DNA.

\begin{figure}[ht!]
  \begin{center}
  \includegraphics[width=7cm]{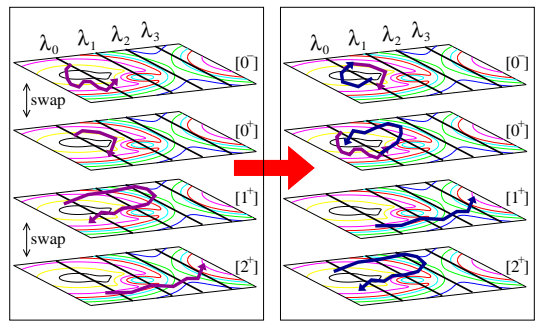}\\
   \caption{(color online) 
Illustration of the swapping move. The picture shows four possible 
pathways on a free energy surface corresponding to the $[0^-],
[0^+], [1^+],[2^+]$ ensembles.  In the next step the swaps
$[0^-] \leftrightarrow [0^+]$ and $[1^+] \leftrightarrow [2^+]$ are performed simultaneously yielding four new pathways. Note that both $[1^+]$ and $[2^+]$
have moved to another reaction channel. The alternative swapping move
$[0^+] \leftrightarrow [1^+]$ would have yielded a rejection as the 
$[0^+]$ does not cross $\lambda_1$.
\label{figswap}}
  \end{center}
\end{figure}

The TIS algorithm works as follows. The first step is to define a RC
and a set of related values $\lambda_0, \lambda_1, \ldots, \lambda_n$
with $\lambda_i < \lambda_{i+1}$. The subsets of phase- or configuration
points for which the RC is exactly equal to $\lambda_i$ basically define
multidimensional surfaces or interfaces. 
These values/interfaces should obey the following requirements:
if the RC is lower than $\lambda_0=\lambda_A$, the system should
be in the reactant
state $A$; if the RC is higher than $\lambda_n=\lambda_B$ the system should be
in the product state $B$; $n$ and the positions for the interfaces in between
should be set to optimize the efficiency.
Furthermore, the surface $\lambda_A$ should be set in such a way that whenever a MD
simulation is released from within the reactant well, this
surface is frequently crossed. The TIS rate expression can then be
formulated as
\begin{align}
\label{kTIS}
k_{AB} &=f_A  {\mathcal P}_A(\lambda_B|\lambda_A)
=f_A \prod_{i=0}^{n-1} {\mathcal P}_A(\lambda_{i+1}|\lambda_i)
  \end{align}
Here, $f_A$ is the flux through the first interface and can be computed by 
straight-forward MD. 
${\mathcal P}_A(\lambda_B|\lambda_A)={\mathcal P}_A(\lambda_n|\lambda_0)$
is the probability that whenever the surface $\lambda_A$ is crossed, 
$\lambda_B$ will be crossed before $\lambda_A$.
The factorization of ${\mathcal P}_A(\lambda_{M}|\lambda_0)$
into
probabilities ${\mathcal P}_A(\lambda_{i+1}|\lambda_i)$ that are much
higher than the overall crossing probability, is the basis of the importance
sampling approach.  
It is important to note that ${\mathcal P}_A(\lambda_{i+1}|\lambda_i)$ are
complicated history dependent conditional probabilities.
If we
consider all possible pathways that start 
at $\lambda_A$ and end by either crossing $\lambda_A$ or $\lambda_B$, while 
having at least one crossing with $\lambda_i$ in between, the fraction that
also crosses $\lambda_{i+1}$ equals 
${\mathcal P}_A(\lambda_{i+1}|\lambda_i)$.
This basically reduces the problem to a correct sampling of trajectories
that should obey the $\lambda_i$ crossing condition.
An effective method to achieve this is the so-called shooting
algorithm~\cite{TPS98_2}.
This Monte Carlo method randomly picks a time slice from the
old existing path and makes a slight modification of this phase point.
Then, this new phase point is used to
propagate forward and backward in time yielding a new trajectory. In TIS, this
propagation is stopped whenever the system enters $A$ or $B$
or, equivalently, whenever $\lambda_0$ or $\lambda_n$ are crossed.
The pathway is then accepted only if the backward trajectory ends in $A$
\emph{and} the total
trajectory  has at least one crossing with $\lambda_i$. In addition,
correct detailed balance rules are applied for the energy and path length 
fluctuations.
The final result follows from the outcomes of a series of 
independent simulations $\{ [{\rm md}],[0^+],[1^+], \ldots, [n^+] \}$. 
The first is the MD
simulation to compute $f_A$. 
The next ones are the path sampling simulations where $[i^+]$ 
indexes the surfaces that has to be crossed.  In the same spirit of
parallel tempering (replica exchange) we could  henceforth try to exchange
the paths from one ensemble to the other, while running all these simulations
simultaneously. The idea was already suggested in~\cite{ErpBol2004} for PPTIS,
but this is the first time that we show its effectiveness for TIS by 
making a small change to the algorithm. In order to have a full flexibility
of swapping moves at all levels, we actually replace the MD simulation
by another path simulation $[0^-]$. This ensemble consists of all possible 
paths that start at $\lambda_A$, then go initially in 
the negative direction and end at the same interface $\lambda_A$.
The flux $f_A$ can now be determined from
\begin{align}
f_A= \Big( \lo t_{\rm path}^{[0^-]} \rc 
+ \lo t_{\rm path}^{[0^+]} \rc \Big)^{-1}
\end{align}
where $\lo t_{\rm path}^{[0^-]} \rc, \lo t_{\rm path}^{[0^+]} \rc$ 
are the average path 
lengths in the $[0^-]$ and $[0^+]$ path ensembles respectively.
The TIS algorithm is then as follows. At each step it is decided 
by an equal probability whether a series of shooting or swapping moves
will be performed. In the first case, all simulations will be updated 
sequentially by one shooting move. In the second case, again an equal 
probability will decide whether the swaps  $[0^-] \leftrightarrow [0^+],
[1^+] \leftrightarrow [2^+], \ldots$ or the swaps $[1^+] \leftrightarrow [2^+],
[3^+] \leftrightarrow [4^+], \ldots$ are performed. Each time that 
$[0^-]$ and $[(n-1)^+]$ do not participate in the swapping move they
are left unchanged. Also when the swapping move does not yield
valid paths for both ensembles, the move is rejected for the two simulations 
and the old paths are counted again. Note that the swapping moves 
do not require any force calculations.
The only exception is $[0^-] \leftrightarrow [0^+]$. Here, the last time step
of the old path in
the $[0^-]$ ensemble is used as initial point to generate a new trajectory
in $[0^+]$ by integrating the equation of motion forward in time. Conversely,
the initial point of the old path in $[0^+]$ is followed backward in time 
to generate a path in $[0^-]$. The two types of swapping moves are illustrated 
in Fig.~\ref{figswap}.

To test the efficiency of this new algorithm, we applied the TIS method with 
and without swapping moves to the denaturation transition of DNA using the
mesoscopic Peyrard-Bishop-Dauxois (PBD) model~\cite{PBD}.   
The PBD
describes the DNA molecule
as an one-dimensional chain of effective atom compounds
yielding the relative base-pair
separations $y_i$ from the ground state positions.
The total potential energy $U$ for an $N$ base-pair DNA chain is then given by
$U(y^N)=V_1(y_1)+\sum_{i=2}^N V_i(y_i) +  W(y_i,y_{i-1})$
with $y^N\equiv \{y_i \}$ the set of relative base pair positions and
\begin{align}
V_i(y_i) &=D_i \Big( e^{-a_i y_i}-1\Big)^2  \label{eqPBD} \\
W(y_i,y_{i-1}) &= \frac{1}{2} K \Big( 1+\rho e^{-\alpha(y_i+y_{i-1})}\Big)(y_i
- y_{i-1})^2 \nonumber
\end{align}
The first term $V_i$ is the Morse potential describing the
hydrogen bond interaction between bases on opposite strands.
$D_i$ and $a_i$ determine the depth and width of this potential
for the  AT and GC base-pairs.
The second term $W$ is the stacking 
interaction. 
All interactions with the solvent and the ions
are effectively included in the force-field. The constants
$K, \rho,\alpha, D_{\rm AT}, D_{\rm GC}, a_{\rm AT}, a_{\rm GC}$
were parameterized in Ref.~\cite{CAGI}. 
For finite chains, the associated state of the molecule is metastable
as would be expected from a double stranded DNA molecule in a 
infinite solution. However, in order to fully dissociate,
all base-pairs should reach the plateau of the Morse potential. As long as one
base pair is still in the stack, it will likely pull all the others back to the
associated state. At ambient conditions, the denaturation is a  
rare event and provides an excellent test for our methods.
As $\lambda(y^N)\equiv \min[\{ y_i \}]$ can describe both 
the associated and the disassociated state, we used this RC. 
For describing the reaction mechanism, a collective variable that 
depends on the positions of all particles might seem more appropriate.
However finding the best possible RC is not the aim of this paper. 
Moreover, the choice of this RC has an additional advantage since 
it allows a very efficient third method that can be used as reference (see below).

We consider a  20 AT base-pair DNA molecule that interacts with a 300 K
Langevin thermostat with  $\gamma=50$ 
ps$^{-1}$ to mimic aqueous damping.
The time step was $\Delta t=1$ fs and base pair masses were 300 amu. 
For the path sampling simulations we used aimless 
shooting~\cite{peters06} where
the velocities of the picked time 
slice are completely
regenerated from Maxwellian distribution. This ensures a higher decorrelation
between accepted paths than the standard shooting move where the velocities
are only slightly changed. As the acceptance rates remained
moderate $\gtrsim 0.3$, the aimless shooting was found to be more efficient for
this system.
After 
a few trial simulations, all interfaces were positioned to have the optimum
${\mathcal P}_A(\lambda_{i+1}|\lambda_i) \approx 0.2$ for all 
$i$~\cite{ErpBol2004,van06}.
After this initialization,
all interface positions were fixed to 
$\lambda_0=0, \lambda_1=0.03, \lambda_2=0.07, \lambda_3=0.13, \lambda_4=0.21,
\lambda_5=0.34, \lambda_6=0.7$, and $\lambda_7=1$ \AA.  
In the next step,
intensive calculations were run for each simulation which consisted 
of $10^9$ simulation steps
for the  MD simulations and $4 \cdot 10^6$ cycles for each path simulation. 
A technique to avoid 
complete separation in the MD simulation was applied~\cite{VanErpPRL}.
Then, we repeated the same simulations with  $2 \cdot 10^6$ cycles using 
a 50 \%  swapping 
probability. The results are shown in table \ref{tabres}. 
\begin{table}[htdp!]
\caption{Results of TIS simulations with and without path swapping. Errors are obtained using block 
averaging.}
\begin{center}
\begin{tabular}{|c|c|c|c|c|}
\hline
& \multicolumn{2}{c|}{Standard TIS} & \multicolumn{2}{c|}{Path Swapping} \\ \cline{2-5}
                          & value  &error(\%)&   value&error(\%) \\ \hline 
          $f_A$(ns$^{-1}$)&   304.9&      5.2&   291.8&      2.2 \\ \hline
${\mathcal P}_A(\lambda_1|\lambda_0)$&   0.256&      9.2&   0.249&      1.2 \\ \hline
${\mathcal P}_A(\lambda_2|\lambda_1)$&   0.244&      4.6&   0.269&      1.9 \\ \hline
${\mathcal P}_A(\lambda_3|\lambda_2)$&   0.246&      5.0&   0.247&      2.4 \\ \hline
${\mathcal P}_A(\lambda_4|\lambda_3)$&   0.310&      2.3&   0.307&      1.8 \\ \hline
${\mathcal P}_A(\lambda_5|\lambda_4)$&   0.310&      1.3&   0.309&      1.2 \\ \hline
${\mathcal P}_A(\lambda_6|\lambda_5)$&   0.210&      1.1&   0.214&      1.3 \\ \hline
${\mathcal P}_A(\lambda_7|\lambda_6)$&   0.533&      0.5&   0.534&      0.8 \\ \hline 
${\mathcal P}_A(\lambda_B|\lambda_A)$&0.000165&     11.5&0.000179&      4.2 \\ \hline
$k_{AB}$ (ns$^{-1}$)      &  0.0492&     12.6&  0.0524&      4.7 \\ 
      2nd     average       &  0.0535&     22.4&  0.0533&       10.3\\ \hline     
\end{tabular}
\end{center}
\label{tabres}
\end{table}

\begin{figure}[ht!]
  \begin{center}
  \includegraphics[width=8cm, angle=0]{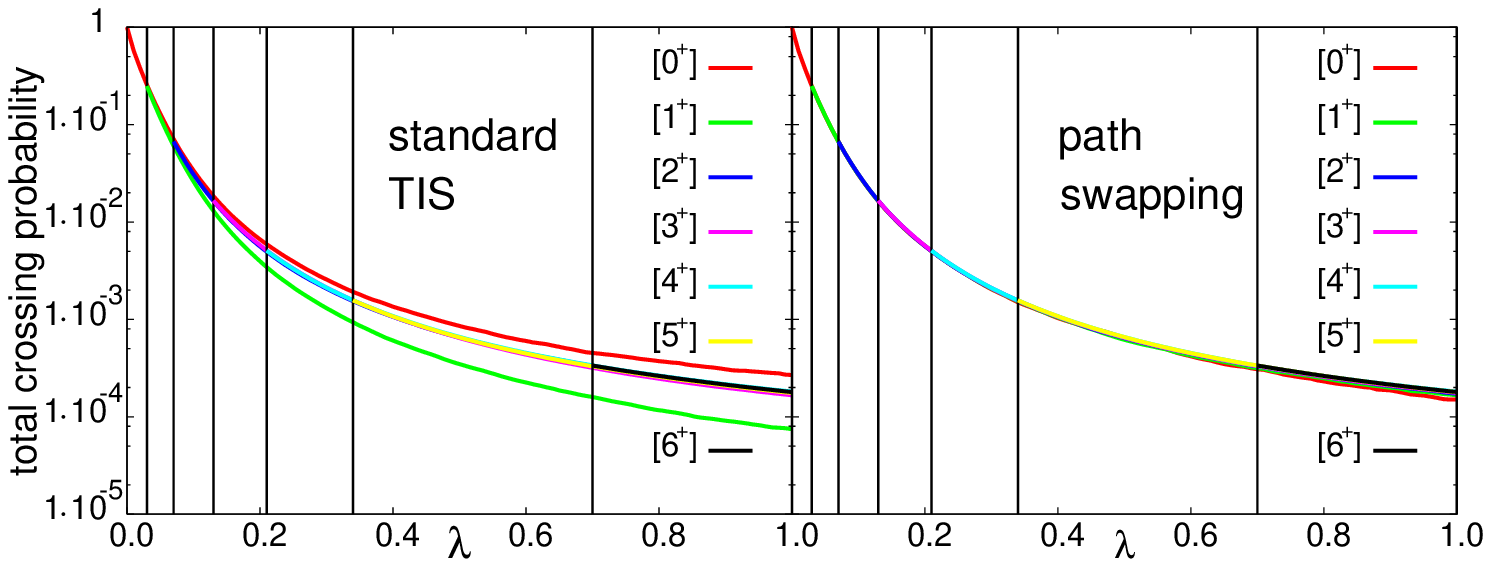} 
   \caption{(color online) The matched overall crossing probabilities for the
   standard TIS and the parallel path swapping. 
\label{OCP}}
  \end{center}
\end{figure}

\begin{figure}[ht!]
  \begin{center}
  \includegraphics[width=8cm, angle=0]{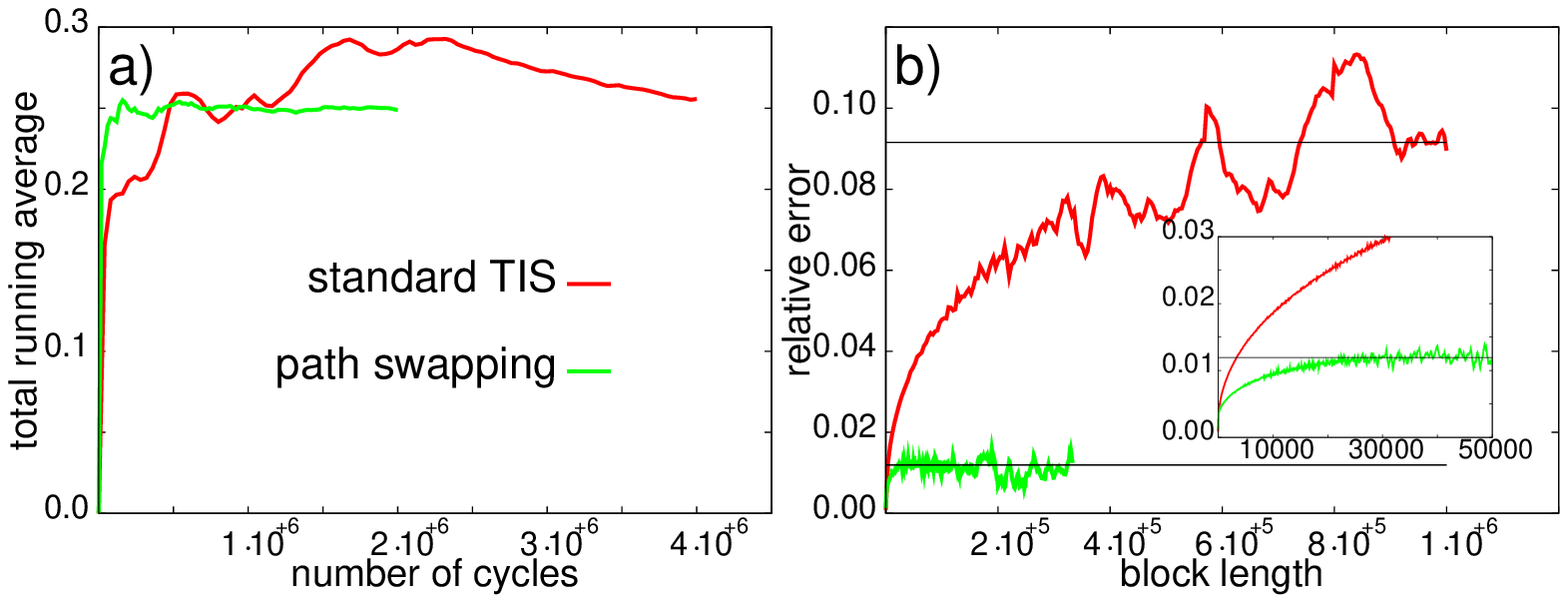}
   \caption{(color online) Error analysis of $[0^+]$ with and without swapping.
   a): the total running average $\sum_i^l x_i/l$ with  $x_i$ the $i$-th measurement. 
b) block average errors $\epsilon(l)=\sqrt{\sum_j (X_j(l) -\bar{x})^2/\bar{x}^2M(M-1)}$ with $X_j$ the 
average of the $j$-th block of length $l$, $\bar{x}$ the total average and $M$ the number of blocks. Horizontal lines at the plateaus indicate the actual relative error $\epsilon$. 
The total correlation is obtained by ${\mathcal N}=(\epsilon/\epsilon(1))^2$.
\label{figerror}}
  \end{center}
\end{figure}

Table \ref{tabres} shows that the parallel path swapping simulations 
have considerable lower errors despite the fewer simulation cycles. 
Also, the construction of the 
overall crossing probabilities in Fig.~\ref{OCP} shows a
much better matching.  
The final error was obtained by the error propagation rule 
$\epsilon_{\rm tot}=\sqrt{\sum \epsilon_i^2}$.
As the validity of these propagation rules is questionable for 
the correlated path swapping simulations, we also calculated the final 
results and errors in a different way. 
Here, we divided the simulations results in five blocks to obtain approximately 
independent rate constants. The errors, 
obtained by the standard deviation, are indeed higher.  
However, the errors 
for standard TIS increased as well by a similar factor. 
Probably, a much higher accuracy is needed to understand the
effect of covariance terms for the errors in the swapping algorithm.
We will come back to these results after we have compared  
the final outcome with that of a third method. This method 
is a very accurate implementation of the RF method
that works due to some special characteristics of the system.
 Like in  standard RF  theory we write
 \begin{align}
 k_{AB}=P_A(\lambda_B) \lo \dot{\lambda}(y^N) \chi[{\rm path}(y^N)] 
\rc_{\{\lambda(y^N)=\lambda_B \}}
 \end{align}
 where $P_A(\lambda_B)$ is the probability that $\lambda(y^N)=\lambda_B$ given that the system
 is in state $A$.  The second 
term is the (unnormalized) transmission coefficient 
 and is calculated by releasing 
 dynamical trajectories starting from the surface $\lambda_B$ and $\chi$ is 
a functional of the trajectory
 which corrects for fast recrossing events.   Although different forms of 
 $\chi$ are used,
 the effective positive flux expression has shown to be the most efficient~\cite{anderson75,van06}.  
 Here $\chi$ is equal to 1 (0 otherwise) 
only if $\dot{\lambda}>0$ and if the backward trajectory crosses
 $\lambda_A$ before $\lambda_B$. 
 The probability $P_A(\lambda_B)$ can be written as
\begin{align} 
P_A(\lambda_B)&=\frac{\int \ud y^N \delta(\lambda(y^N)-\lambda_B) 
e^{-\beta U(y^N)}}{\int \ud y^N  \theta(\lambda_B-\lambda(y^N))
e^{-\beta U(y^N)}} \label{PA} \\
&=
\frac{\sum_i \int \ud y^N \delta(\lambda(y_i)-\lambda_B)
\prod_{j \neq i} \theta(y_j - \lambda_B)
e^{-\beta U(y^N)}}{\int \ud y^N
(1-\prod_k \theta(y_k-\lambda_B) ) e^{-\beta U(y^N)}} \nonumber
\end{align}
As the integrals of Eq.~(\ref{PA})  are all of a special factorial form, we can apply the direct integration
method of~[\onlinecite{VanErpPRL}] yielding a precision of several digits.
In the next step, we need to generate a representative set of configurations
on the surface $\lambda_B$. 
It can be shown that ensemble averages with potential $U(y^N)$ and 
fixed $\min[\{y_i\}]=\lambda_B$, are in fact identical to that of a 
freely moving chain on a translational invariant potential $U'$ that is 
related to $U$ by  
$U'(y^N) \equiv U( y^N-\lambda(y^N)+\lambda_B)$.
Hence, 
we can generate the required surface points by running a MD simulation using 
$U'(y^N)$, save every 1000th time step to dissolve correlations, and shift 
these configurations to the surface $\lambda_B$.  From these points, we
release  trajectories using normal potential $U$ and calculate $\chi$. 
We applied this method using a numerical integration step of $dy=0.01$~\AA  
~yielding the 
result $P_A(\lambda_B)=5.316 \cdot 10^{-3}$ \AA$^{-1}$.
Then we released $4 \cdot 10^6$ trajectories for which the initial points 
were generated 
using the dynamical shifted potential $U'(y^N)$. The transmission
coefficient yielded $9.854 \pm 0.066$  \AA/ns and the combined result  
$k_{AB}=0.0524$ ns$^{-1}$,
which is in excellent agreement with the TIS results. 

Now we come back to the results of table~\ref{tabres} and try to express 
the efficiency into the 
so-called efficiency times $\tau_{\rm eff}$ that are defined as the number 
of force calculations
required to obtain a statistical error equal to 1. For the simulations 
$[0^+],[1^+],\ldots$ the 
efficiency times are given by~\cite{van06}
\begin{align}
\tau_{\rm eff}^{[i^+]}=\frac{1-p_i}{p_i} \xi_i L_i {\mathcal N}_i
\end{align} 
Here, $p_i={\mathcal P}_A(\lambda_{i+1}|\lambda_i)$ 
and $L_i=\lo t_{\rm path}^{[i+]} \rc/\Delta t$ which are in principle 
independent from the simulation method. $\xi_i$ is the ratio between the average 
cost of a 
simulation cycle and $L_i$.  ${\mathcal N}_i$ is the effective correlation. 
\begin{table}[hbdp]
\caption{Efficiency analysis}
\begin{center}
\begin{tabular}{|c|c|c|c|c|c|c|c|c|}
\hline 
& \multicolumn{4}{c|}{Standard TIS} & \multicolumn{4}{c|}{Path Swapping} \\
\cline{2-9}
                &$L$&$\xi$&${\mathcal N}$&$\tau_{\rm eff}(10^5)$ & $L$&$\xi$&${\mathcal N}$&$\tau_{\rm eff}(10^5)$\\ \hline
$[md]/[0^-]$    &    1&1    &             824&                          27 &  3262& 0.87&             146&                          27\\ \hline
$[0^+]$         &  115&1.16 &           11519&                          45 &   108& 0.83&              95&                         0.3\\ \hline
$[1^+]$         &  289&1.05 &            2087&                          20 &   325& 0.53&             261&                           1\\ \hline
$[2^+]$         &  764&1.01 &            3284&                          78 &   765& 0.51&             377&                           4\\ \hline
$[3^+]$         & 1832&0.98 &             921&                          37 &  1827& 0.49&             272&                           5\\ \hline
$[4^+]$         & 3768&0.94 &             327&                          26 &  3776& 0.47&             139&                           6\\ \hline
$[5^+]$         & 7464&0.87 &             121&                          29 &  7483& 0.43&              97&                          11\\ \hline
$[6^+]$         &14391&0.70 &             109&                          10 & 14340& 0.35&             147&                           6\\ \hline
overall         &     &     &                &                       14907 &      &     &                &                         648\\ 
2nd aver.      &     &     &                &                       46660 &      &     &                &                        2486\\ \hline
\end{tabular}
\end{center}
\label{tabeff}
\end{table}
The results of the efficiency analysis
are given in table ~\ref{tabeff} and  show that the swapping   moves decrease 
both $\xi$ and ${\mathcal N}$. The efficiency times are all lowered by at 
least a factor
of 2 for all path simulations. 
Spectacular is the decrease of correlation from 11519 to 95 in the $[0^+]$ 
simulation 
yielding an increase in efficiency
of a factor $150$. Inspection of  Fig.~\ref{figerror}-a) reveals large 
fluctuation in the overall running average for standard TIS 
even after $4\cdot 10^6$ cycles. In contrast, the swapping results 
shows a much faster convergence. This is also reflected
in the block-error analysis of Fig.~\ref{figerror}-b). 

The overall efficiency time is derived from $\tau_{\rm eff}=\epsilon^2 \tau_{\rm sim}$.
Here $\epsilon$ the relative error in $k_{AB}$ and $\tau_{\rm sim}$ is the total simulation time.
The two ways of averaging show that the swapping moves give an overall improvement of  approximately 20. However, it is important to realize
that applying the same number of cycles for each simulation does not give the best possible performance.
From the results 
of~[\onlinecite{van06}], one can show that the efficiency can be improved by a factor of 7 if the 
optimal ratio of cycles proportional to $\propto \sqrt{\tau_{\rm eff}^{[i^+]}}/\xi_i L_i$ is  applied.
However, we believe that the path swapping efficiency can be improved by a 
similar factor if we change the algorithm 
to allow an unequal 
distribution of shooting moves among the simulations. We are now working on such algorithms.

To conclude, we have shown that TIS combined with path swapping can give a huge improvement of efficiency. 
For the denaturation of 
the PBD model of DNA, we obtained an improvement of approximately a factor 20. Individual path 
simulations were improved upto two
orders of magnitude. 
Therefore, we believe that parallel path swapping can become an important method in any type of
rare event simulations.

I would like to thank Paolo Pescarmona for carefully reading this paper.

\bibliographystyle{prsty}

\end{document}